\documentclass{aa}  

\usepackage{graphicx}
\usepackage{txfonts}
\usepackage{booktabs}
\usepackage{arydshln}
\usepackage[nodayofweek,level]{datetime}
\usepackage{threeparttable}
\usepackage{orcidlink}

\newcommand{\orcid}[1]{\href{https://orcid.org/#1}{\textcolor[HTML]{A6CE39}{\aiOrcid}}}
\newcommand{\pl}{HD~332231~b\xspace}
\newcommand{\host}{HD~332231\xspace}
\newcommand{\transitnight}{4 August, 2020\xspace}
\newcommand{\fref}[1]{Fig.~\ref{#1}}
\newcommand{\tref}[1]{Table~\ref{#1}}
\newcommand{\sref}[1]{Section~\ref{#1}}

\usepackage[normalem]{ulem}
\usepackage{color}

\begin{document}

    \title{Orbital alignment of HD~332231~b}

   \subtitle{The warm Saturn HD~332231~b/TOI-1456~b travels on a well-aligned, circular orbit around a bright F8 dwarf}

   \author{E.~Knudstrup 
          \inst{1,2}\orcidlink{0000-0001-7880-594X}
          \and
          S.~H.~Albrecht\inst{1}\orcidlink{0000-0003-1762-8235}
          }

   \institute{Stellar Astrophysics Centre, Department of Physics and Astronomy, Aarhus University, Ny Munkegade 120, DK-8000 Aarhus C, Denmark\\
              \email{emil@phys.au.dk}
         \and
             Nordic Optical Telescope, Rambla José Ana Fernández Pérez 7, E-38711 Breña Baja, Spain\\
             }

   \date{Received September 15, 1996; accepted March 16, 1997}

 
  \abstract
   {
   Contrary to the orthodox picture of planet formation resulting in a neatly ordered Solar System, exoplanet systems exhibit highly diverse orbits: short and long periods, circular and eccentric, well- and misaligned, and even retrograde orbits. In order to understand this diversity it is essential to probe key orbital parameters. Spin--orbit alignment is such a parameter and can provide information about the formation and migration history of the system. However, tidal circularisation and alignment might hamper interpretations of orbital eccentricity and obliquities in the context of planet formation and evolution for planets on orbits shorter than about 10 days. 
  }
   {
   Here we aim to measure the projected stellar obliquity in the \host system in which a warm (period$\approx18.7$~days) giant planet orbits a bright F star on a circular orbit.
   }
   {We observed the system during a transit with the HARPS-N spectrograph and obtained data on the Rossiter-McLaughlin effect. We analysed the spectroscopic transit data together with new TESS photometry employing three different analysis methods. 
   }
   {The results from the different approaches are fully consistent. We find a projected obliquity of $-2\pm6$~$^{\circ}$, indicating the stellar spin axis to be well-aligned with the orbit of the planet. We furthermore find evidence for transit timing variations suggesting the presence of an additional third body in the system. 
  }
   {Together with the low orbital eccentricity, the good alignment suggests that this warm giant planet has not undergone high-eccentricity migration.  
   }


   \maketitle
%

\section{Introduction}

Our understanding of planet formation is intimately linked to our knowledge of the migratory patterns of giant planets. The formation of so-called hot Jupiters (HJs), gas giant planets with orbits of less than some $10$ days, and warm Jupiters (WJs) found at larger separations with orbital periods of between $\approx$ 10 and 200~days is not well understood \citep[see][for a review]{art:dawson2018}. In situ formation appears unlikely, at least for the inner planets. Therefore, orbital migration from their original birth orbits to the orbits we now observe them in appears to be an attractive explanation. However, the exact route(s) for such migration remains poorly understood. 

The two leading theories for orbital shrinkage are ``disc-migration'' and ``high-eccentricity (high-$e$) migration'' \citep{art:dawson2018}. In disc-migration, angular momentum is exchanged between the planet and the planetary disc, which leads to in-spiraling of the planet \citep[e.g.][]{art:lin1996,art:Baruteau+2014}. High-$e$ migration is the result of interactions between multiple bodies in the planetary system. Here the migration can be caused by scattering, which creates a highly eccentric orbit that subsequently shrinks via tidal circularisation \citep[e.g.][]{art:nagasawa2008,chatterjee2008}, or by a distant stellar or planetary companion in the system, which causes secular Kozai Lidov cycles followed by tidal friction \citep[e.g.][]{wu2003,art:fabrycky2007,naoz2016}. In systems with three or more planets, exchange of angular momentum can drive the Jupiter's orbit to large eccentricities in a process known as secular chaos \citep[e.g.][]{Laskar2008,art:wu2011,teyssandier_lai_vick2019}, and subsequent tidal orbital shrinking can lead to a close-in orbiting giant planet.

In general, multi-body interactions in ``high-$e$ migration'' perturb the original orbit of the planet, leading to an elliptical or eccentric and inclined orbit with respect to its original orbital plane. Conversely, disc-migration is expected  to result in low-eccentricity near-circular orbits located near the midplane of the protoplanetary disc in which the planet formed. Assuming alignment between the stellar equator and the protoplanetary disc, stellar obliquities (the angle between the orbital and stellar angular momenta) can be used next to eccentricity measurements to inform theories about planet formation and evolution \citep[e.g.][]{fabrycky_winn2009,art:triaud2010,art:dawson2018}. 

However, such inference is complicated by a number of factors. It has been found that tidal interactions can significantly alter both the obliquity \citep{winn2010,AlbrechtWinnJohnson+2012} and the eccentricity \citep[e.g.][]{husnoo2012,art:bonomo2017}. Furthermore, good alignment between the protoplanetary disc and stellar equator is not guaranteed, despite the fact that the former inherits its angular momentum from the latter. Chaotic accretion might lead to a misaligned stellar and protoplanetary disc spin as suggested by for example  \citet{bate2010,Thies+2011,fielding2015,bate2018}; but see also \citet{Takaishi_Tsukamoto_Suto_2020} who suggest that moderate misalignment can also be created that way. Magnetic torques might also lead to misalignment \citep{foucart2011,lai2011,Romanova2021MNRAS}. Inclined stellar or planetary companions might tilt discs \citep[see, e.g.][]{Borderies_Goldreich_Tremaine1984,Ludow_Ogilvie2000,Batygin2012,spalding2014,matsakos2017}, although \citet{ZanazziLai2018} found that HJs suppress such misalignment. While most systems observed so far suggest good primordial alignment \citep{art:albrecht2013} at least one system appears to have had a retrograde spinning protoplanetary disc \citep{Hjorth+2021}. Finally, orbits might have large inclinations relative to the stellar spin, but this is the result of precession caused by a giant orbiting a planetary companion on a wide orbit \citep{Huber+2013,gratia2017}. 

\begin{figure*} 
        \includegraphics[width=\textwidth]{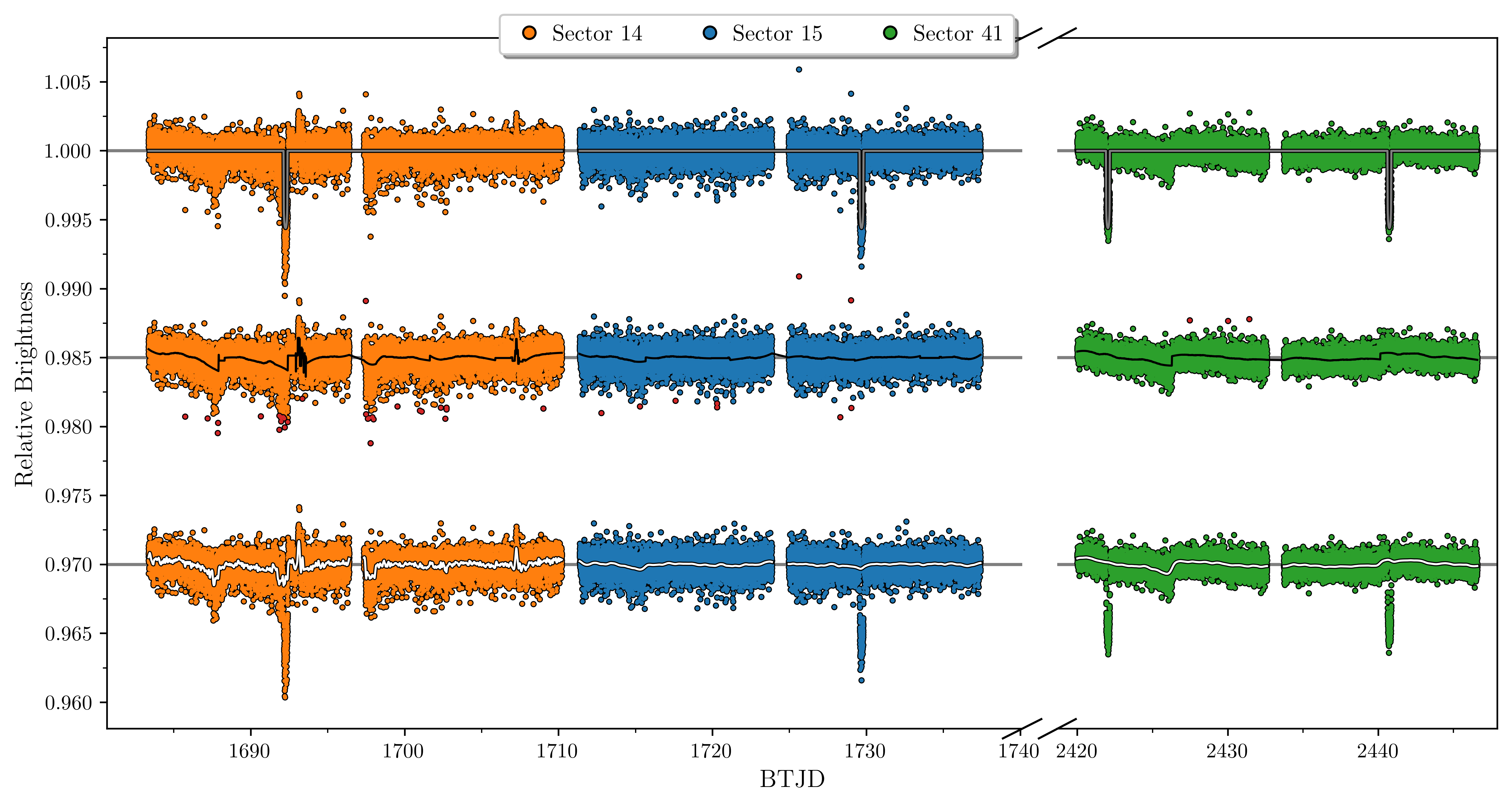}
    \caption{{\bf TESS light curve for HD~332231.} The top curve is a normalised but not detrended light curve of HD~332231 as observed by TESS (top curve), with orange, blue, and green points corresponding to Sectors 14, 15, and 41, respectively. The grey line is a transit model created from the parameters in \citet{art:dalba2020}. The transit model has been used to temporarily remove the transit in the light curve offset by -0.015 (denoted as the straight line below the points). Here the black line shows a Savitsky-Golay filter \citep[as implemented in][]{misc:lightkurve} used to filter and detrend the data. The red points are outliers removed through a 5$\sigma$ sigma clipping. 
    The light curve with outliers removed and the transits re-injected is shown in the light curve offset by -0.03, where the white line shows the GP (see~\sref{sec:analysis}) used for detrending.}
    \label{fig:TESS}
\end{figure*}

One class of systems suited to probing the evolution pathways of giant planets is the WJs, as tides may not have altered the obliquity, nor completely dampened the eccentricity. Depending on the presence or absence of a nearby stellar or planetary companion, a number of formation and evolution processes may be excluded or considered. 

\host is one such system. It was first detected by the Transiting Exoplanet Survey Satellite \citep[TESS;][]{art:ricker2015} and was given the ID TOI-1456 \citep[TOI: TESS Object of Interest;][]{art:guerrero2021}. It was subsequently confirmed and characterised by \citet{art:dalba2020} through radial velocity (RV) measurements. With an orbital period of $\sim 18.7$~d, it was determined to be a WJ, with a radius slightly larger than that of Saturn ($R_\mathrm{p} \sim 0.87$~R$_\mathrm{Jupiter}$), but with a significantly lower mass ($M_\mathrm{p} \sim 0.24$~M$_\mathrm{Jupiter}$). Key parameters for the \host system are summarised in \tref{tab:toi-1456_para}.
 
Here we want to gain information on the obliquity of the host star in order to further investigate its history. We observed the system with a high-resolution spectrograph while the planet was transiting its host. Analysis of the Rossiter-McLaughlin \citep[RM;][]{art:rossiter1924,art:mclaughlin1924} effect \cite[e.g.][]{Struve1931,art:queloz2000,winn2005,art:albrecht2007,hebrard2008}, which is a line shape distortion occurring during transits, allows us to determine $\lambda$, the sky projection of the stellar obliquity, $\psi$.

We describe our observations in \sref{sec:data}. In \sref{sec:analysis} we use three approaches to determine the projected obliquities from the obtained data sets. After describing our main results in \sref{sec:results}, we briefly discuss them in the context of measurements for similar systems in \sref{sec:discussion} before presenting our conclusions.
 
\section{Observations}
\label{sec:data}

\host was observed by TESS in Sector 14 and 15 with a single transit in both sectors. An additional transit occurred in the observational gap between these two sectors. As noted by \citet{art:dalba2020}, the transit in Sector 14 was heavily affected by scattered light and was masked out by the presearch data conditioning (PDC) module in the Science Processing Operations Center \citep[SPOC;][]{art:jenkins2016}. We include all data and correct for this extra background using the \texttt{RegressionCorrector} implemented in \texttt{lightkurve} \citep{misc:lightkurve}. The background-corrected and normalised light curve is shown in \fref{fig:TESS}. HD 332231 was observed again in the extended mission of TESS in Sector 41 with two consecutive transits. This allows improved determination of the ephemeris as well as other photometric transit parameters, especially as the scatter in the Sector 41 light curve is significantly lower than that in Sectors 14 and 15 as seen in \fref{fig:TESS}.

Despite correcting for the background, there are still some outliers in the light curve. In an attempt to remove these outliers, we firstly removed the transits using the best-fitting parameters from Dalba et al. (2020) as shown by the grey line in the top curve of \fref{fig:TESS}. We then applied a Savitsky-Golay filter \citep{art:savitzky1964}, as implemented in \texttt{lightkurve}, which is shown as the black line in the middle curve of Fig. 1. Finally, we removed the outliers shown as red points in the middle curve of \fref{fig:TESS} through $5\sigma$ sigma clipping. After having background-corrected and cleaned the light curve, we re-injected the transit, which can be seen in the bottom curve.

\begin{table}
\centering
\caption{Parameters of the HD\,332231 system.} \label{tab:toi-1456_para}

\begin{threeparttable}
\begin{tabular}{lc}
\toprule
 Parameter & {Value}\\ 
\midrule
Alternative name\tnote{a}     & TYC 2689-70-1\\
Alternative name\tnote{b}              & TOI-1456\\
R.A. (J2000)                     & 20:26:57.92 \\
Dec. (J2000)                    & +33:44:40.02 \\
Parallax (mas)\tnote{c}                  & $12.37 \pm 0.03$ \\
$V$ magnitude\tnote{a}          & 8.56$\pm$0.01 \\
Effective temperature (K)\tnote{d} & 6089$^{+97}_{-96}$  \\
Surface gravity (dex)\tnote{d}              & 4.279$^{+0.027}_{-0.034}$  \\
Metallicity (dex)\tnote{d}           & 0.036$^{+0.059}_{-0.058}$  \\
Stellar mass (M$_\odot$)\tnote{d}              & 1.127$\pm 0.077$   \\
Stellar radius (R$_\odot$)\tnote{d}           & 1.277$^{+0.039}_{-0.036}$    \\
Age (Gyr)\tnote{d} & $4.3^{+2.5}_{-1.9}$\\
Period (days)\tnote{d}              & $18.71204 \pm 0.00043$   \\
Eccentricity\tnote{d}           &   $ 0.032^{+0.030}_{-0.022}$    \\
Planetary mass (M$_{\rm Jupiter}$)\tnote{d}              & $ 0.244\pm 0.021 $   \\
Planetary radius (R$_{\rm Jupiter}$)\tnote{d}           &   $0.867^{+0.027}_{-0.025}$    \\

\bottomrule
\end{tabular}
\begin{tablenotes}
    \item[a] \citet{art:hog2000}.
    \item[b] \citet{art:guerrero2021}.
    \item[c] \citet{art:gaia2018}.
    \item[d] \citet{art:dalba2020}.
\end{tablenotes}
\end{threeparttable}
\end{table}

\begin{figure}
    \centering
    \includegraphics[width=\columnwidth]{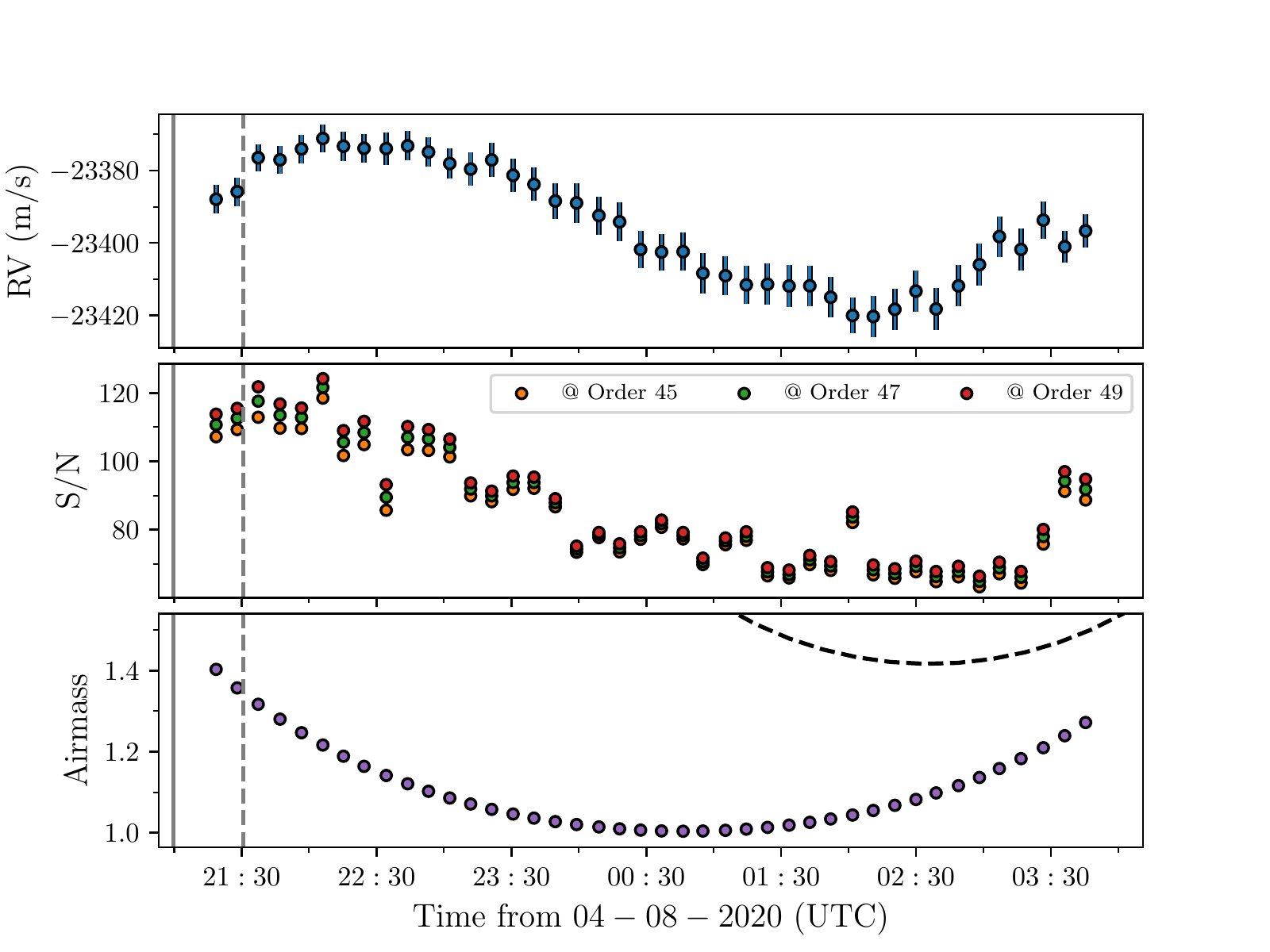}
    \caption{{\bf The HARPS-N data on the transit night.} {\bf Top:} RVs extracted from the HARPS-N pipeline plotted against time. {\bf Middle:} S/N for each exposure at three different orders, namely 45, 47, and 49. {\bf Bottom:} Airmass for each exposure plotted with the airmass of the Moon plotted as the dashed line. In each panel, the vertical solid and dotted lines denote nautical and astronomical twilight, respectively.}
    \label{fig:transitnight}
\end{figure}

To determine the projected obliquity in \host, we obtained spectroscopic transit data with the High Accuracy Radial velocity Planetary Searcher North \citep[HARPS-N;][]{art:mayor2003,art:cosentino2012} mounted on the 3.58~m Telescopio Nazionale Galileo (TNG) located on Roque de los Muchachos, La Palma, Spain. We observed a transit occurring during the night of \transitnight, with observations starting at 21:05 UT until 03:40 UT (programme ID: A41/TAC19, P.I.~Knudstrup). The exposure time was set to 540~s and with an overhead of roughly 20~s, the sampling was approximately 560~s. For comparison, the total transit lasts about $6.1$~h. The RVs and $1\sigma$ uncertainties obtained through the Data Reduction Software (DRS) of HARPS-N are printed in \tref{tab:RVs} and shown in the top panel of \fref{fig:transitnight}. The middle panel shows the signal-to-noise ratio (S/N) for each exposure for three orders, and the airmass, ranging from 1.4 to 1.0, is plotted in the lower panel along with the airmass of the Moon plotted as a dashed line. The seeing was variable with values between 0.9 and 2.0 arcsec with a median of 1.5 arcsec. 

\begin{figure}
        \includegraphics[width=\columnwidth]{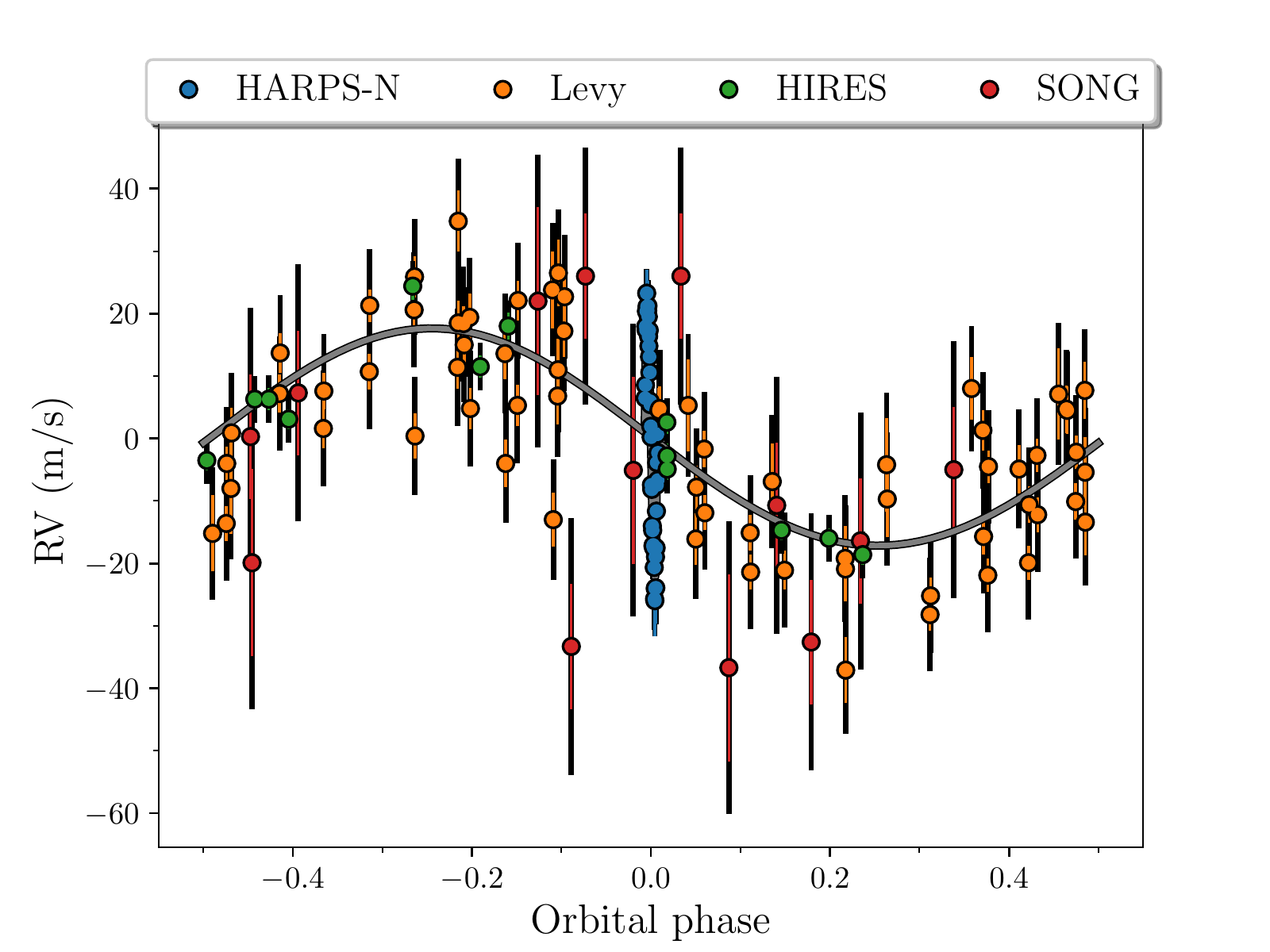}
    \caption{{\bf Radial velocity curve for HD~332231~b.} Radial velocities from HARPS-N, Levy, HIRES, and SONG shown with blue, orange, green, and red error bars, respectively. The grey line is the best-fitting model of the orbit modulated by the RM effect, which is obscured by the HARPS-N data. The coloured error bars are the nominal errors, and the black error bar is the nominal error with the jitter term added in quadrature. }
    \label{fig:rv}
\end{figure}

We supplement our HARPS-N transit observations with the RVs presented in \citet{art:dalba2020}. These include RVs obtained using the Levy Spectrograph \citep{art:radovan2010} at the Automated Planet Finder \citep[APF;][]{art:radovan2014,art:vogt2014}, the High Resolution Echelle Spectrometer \citep[HIRES;][]{art:vogt1994} at the Keck I telescope, and the Hertzsprung node of the Stellar Observations Network Group \citep[SONG;][]{art:andersen2014,art:grundahl2017}. The RVs from \citet{art:dalba2020} as well as our HARPS-N observations are shown in \fref{fig:rv}.

\section{Determining the projected stellar obliquity in the \host system}
\label{sec:analysis}

When part of the rotating stellar surface is blocked from view, the rotational broadened stellar absorption lines are distorted relative to their uneclipsed shape. The distortion and its time evolution are governed by the projection of the angle between the stellar spin axis and the orbital angular momentum of the occulting body. 

Using the HARPS-N data from the night on the \transitnight, supplemented by the TESS photometry and the publicly available RV data described above, we measure $\lambda$ employing three different approaches. We are motivated to do so as different analysis methods can have different dependencies on systematic errors towards some system parameters; for example, timing offsets and orbital inclination. We used the following approaches: i) We analysed the stellar absorption lines and distortions thereof during the transits themselves. For this, we used the cross-correlation function (CCF), which serves as an "average stellar absorption line" as delivered by the DRS. ii) We also employed a procedure where we first measured the position of the CCF distortion in RV space, and then used these subplanetary velocities, $v_p$, to determine the projected obliquity. iii) Finally, we determined the projection of $\psi$ using the anomalous stellar RVs occurring during transit as a result of the line distortions. As all three methods use the same data sets, they are expected to deliver fully consistent results. 

Before describing the specific steps for each approach, we outline the setup common to all three approaches. In each case, we created a model which we compared to the data sets, the spectroscopic transit data, TESS photometry, and RVs taken outside transits. We extracted the confidence intervals of relevant parameters using a Markov Chain Monte Carlo (MCMC) as described below.

Parameters mainly governed by photometric data are the orbital period, $P$, a specific mid-transit point, $T_0$, the planet-to-star radius ratio, $r/R$, the scaled semi-major axis, $a/R$, the cosine of the orbital inclination, $\cos i_\mathrm{o}$, and the quadratic limb-darkening parameters, $q_1$ and $q_2$. For all three approaches to determining the projected obliquity, we modelled the TESS light curves with the \citet{art:mandel2002} formalism as implemented in the \texttt{batman} package \citep{art:kreidberg2015}. As is evident from the bottom panel of \fref{fig:TESS} the light curves are affected by systematic error. Therefore, at each step of our MCMC analyses, we adjusted a Gaussian Process (GP) model to the TESS data, where we modelled the correlated noise with a Matérn-3/2 kernel implemented in \texttt{celerite} \citet{celerite}. As the systematic error is quite different between the three sectors, we include three sets (one for each sector) of the two hyper parameters; the amplitude of the noise, $A$, and the timescale, $\tau$. We furthermore include a jitter term for each light curve, $\sigma_{\mathrm{Sector}\ i}$ ($i=$ 14, 15, 41).

For reasons discussed in \sref{sec:results} we also included one additional parameter, $\Delta T_0$, here in our final model to data comparison. This parameter allows the spectroscopic transit midpoint to float with respect to the midpoint given by the linear ephemeris ($P,T_0$). 

Parameters mainly determined using out-of-transit RVs are the orbital RV semi-amplitude, $K$, the orbital eccentricity, $e$, the argument of periastron, $\omega$, and the RV offsets for the four spectrographs, $\gamma_i$ ($i=$ HARPS-N, Levy, HIRES, SONG), as well as their RV jitter terms added in quadrature to the RV uncertainties provided by the RV pipelines $\sigma_{i}$. Parameters describing the shape of the CCFs and any distortions during transits are the projected stellar rotation speed, $v \sin i$, the stellar surface motion, which, here, is parametrized by macro-turbulence, $\zeta$, and micro-turbulence, $\xi$ \citep{gray2005} . The HARPS-N instrument provides spectra with a resolution of $R\approx115\,000$ resulting in a spectral point spread function (PSF) with a FWHM of $\sim2.6$~km\,s$^{-1}$ or a $\sigma_{\rm PSF}$ of $\sim1.1$~km\,s$^{-1}$. We use this value to create a Gaussian with which we convolve our model CCFs. In addition, we require two parameters with which we attempt to capture the limb darkening in the band pass of HARPS-N, $q_{1,{\rm HARPS-N}} $ and $q_{2,{\rm HARPS-N}}$. The deformation of the lines during the transit in our model is chiefly governed by the projected obliquity, $\lambda$ (our parameter of main interest), the above-mentioned $v \sin i$, and the impact parameter $b \equiv a/R \cos i_\mathrm{o}$, which is also controlled by the photometric data obtained during transit.

To obtain confidence intervals for our system parameters, in addition
to the data described in \sref{sec:data}, we employed  prior information on stellar limb-darkening and stellar surface fields. With the stellar parameters determined by \citet{art:dalba2020} and listed in \tref{tab:toi-1456_para}, we queried the tables provided by \citet{art:claret2013} and \citet{art:claret2017} for values for quadratic limb-darkening parameters. For the TESS passband, we find $q_1=0.253$ and $q_2=0.289$, and for the $V$ band used for our spectroscopic transit data we obtain $q_{1,{\rm HARPS-N}}=0.513$ and $q_{2,{\rm HARPS-N}}=0.199$. From $T_\mathrm{eff}$, $\log g$, and the relationship presented in \citet{art:doyle2014}, we find $\zeta=4.46$~km\,s$^{-1}$. Assuming a sigma of $1$~km\,s$^{-1}$ we use this as a Gaussian prior. For the micro-turbulence parameter $\xi$ we assume a Gaussian prior with a mean of $2$~km\,s$^{-1}$ and a $\sigma$ of $1$~km\,s$^{-1}$, which is in line with the value given in \citet{art:hirano2011}. All parameters and their priors can be found in \tref{tab:mcmc_res}.

\begin{figure}
        \includegraphics[width=\columnwidth]{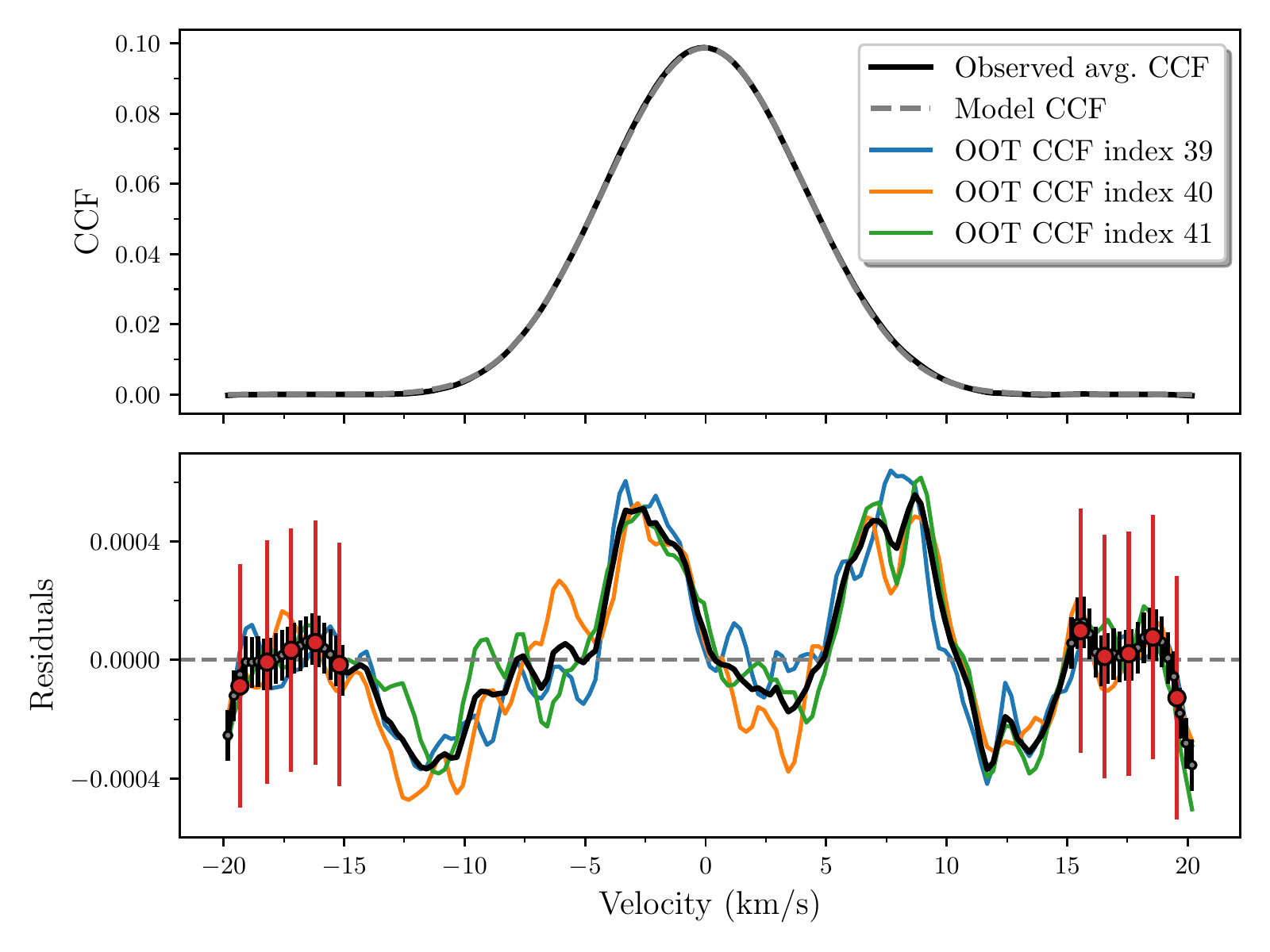}
    \caption{{\bf Out-of-transit cross-correlated-functions.} {\bf Top:} Average observed CCF (created from the
three OOT CCFs) shown as the solid black line with the best-fitting model overlain as the grey dashed line. The  surface area under the CCF was arbitrarily set to 1. All CCFs have been shifted into the stellar rest frame. {\bf Bottom:} Residuals between the OOT CCF and model. We also show the three epochs used to create our average out of transit CCF, i.e. the three last observations on the \transitnight (Indexing started with 0). Shown as grey error bars is the scatter on the CCFs outside of the central peak, specifically $\pm 15$~km~s$^{-1}$, which we have binned to 1~km~s$^{-1}$ and scaled to get a $\chi^2_\nu \sim$1. The binned and scaled errors are shown with red error bars, which are applied to all data points, but for illustrative purposes are only shown in the aforementioned range.} 
    \label{fig:ccf}
\end{figure}

Before proceeding with creating a model and comparing it to our data, two additional steps concerning the HARPS-N CCFs were required. As mentioned above, the PSF of HARPS-N has a $\sigma_{\rm PSF}$ of $\sim1.1$~km\,s$^{-1}$. The CCFs provided by the DRS are sampled on a velocity grid with a bin size of $0.25$~km\,s$^{-1}$. We therefore binned the CCFs onto a velocity scale with bins of $1$~km\,s$^{-1}$ in width. Secondly, to assign the proper weight to each type of data (photometry, out-of-transit RVs, HARPS-N transit night data), the different data sets require properly scaled uncertainties. To obtain such uncertainties for the CCFs, we performed a fit to the last three CCFs obtained during the transit night, as these were obtained after egress. The averaged out-of-transit (OOT) CCF was fitted with the relevant parameters from above ($v \sin i$, $\zeta$, $\xi$, and limb-darkening). From this best fit, and requiring a reduced chi-squared, $\chi_\nu^2$, to be $\approx1,$ we obtain uncertainties of $\sim 0.0004$ for each velocity point in the CCF. Furthermore, we normalised our CCFs by setting the surface area under the CCF to 1 for both data and model. In \fref{fig:ccf} we display the OOT CCFs, the best fitting model, and the residuals of the CCFs. At this point, we were in a position to compare the data to the different models.

\subsection{Line distortions: planet shadow}
\label{sec:line_distortions}

Here we present our analysis of the deformations of the stellar lines directly caused by the planetary transits. To perform such an analysis, we followed the approach by \citet{art:albrecht2007,art:albrecht2013,Hjorth+2021}, but see also for example \citet{Brown+2012,johnson2014,Zhou+2016}. Our model CCFs were created by constructing a limb-darkened stellar grid, where we assume a quadratic limb-darkening law of the form
\begin{equation}
\label{equ:limb_darkening}
    I = 1-u_1(1-\mu)-u_2(1-\mu)^2\,\,\,.
\end{equation}
Here $\mu = \cos(\theta)$ with $\theta$ being the angle between the local normal and a line parallel to the line of sight, and $I$ is the local intensity normalized by the intensity at the center of the disc, i.e., $\mu=1$.

For each pixel of the stellar surface, we include the effects of macro- and micro-turbulence following \citet{gray2005}. The local line profile in our model is given by
\begin{equation}
    \Theta(v) = \frac{1}{2\sqrt{\pi}\zeta} 
     \left( \frac{\exp(-v/\zeta\cos\theta)^2}{\cos\theta} + \frac{\exp(-v/\zeta\sin\theta)^2}{\sin\theta}\right) \, ,
\end{equation}
where we assume equal velocities and surface areas for tangential and radial flows.

Assuming solid body rotation (no differential surface rotation), the radial velocity of the stellar surface is a function of the distance from the stellar spin axis only. We tested whether or not including differential rotation into our model would change the main conclusion we draw about the projected obliquity and found this not to be the case. This is because differential rotation is difficult to measure via the RM effect \citep{art:hirano2011} and would require a very well suited data set. In our case in particular, the planet covers only a very small range in stellar latitude, and so even strong differential rotation would not change the shape of the RM signal. Still, in the presence of strong differential surface rotation, the disc-integrated $v \sin i$ might differ from the "local" $v \sin i$ as probed by $v_p$, but given the data at hand, this would not be noticeable.

We define a coordinate system so that the $x$-axis is oriented along the stellar equator and the $y$-axis parallel to the projected stellar spin axis. The Doppler velocity of the stellar surface below a planet $v_\mathrm{p}$ is then simply given by the distance from the y-axis and the projected stellar rotation speed,
\begin{equation}
\label{equ:v_p}
    v_p = \frac{x}{R} v \sin i \, . 
\end{equation}

The position of the planet $(x_\mathrm{p},y_\mathrm{p})$ in this coordinate system is given by
\begin{equation}
 \begin{pmatrix} x_\mathrm{p} \\ y_\mathrm{p} \end{pmatrix} = 
 \begin{pmatrix} - \frac{\Gamma}{R} \cos(\omega+\nu)  \\ - \frac{\Gamma}{R} \sin(\omega+\nu) \cos i_o \end{pmatrix}
  \begin{pmatrix} \cos \lambda & -\sin \lambda \\ \sin \lambda & \cos \lambda \end{pmatrix} \, .
\end{equation}
Here, $\nu$ represents the true anomaly and $\Gamma$ the distance between the centre of the planet and that of the star on the Keplerian orbit.\footnote{Normally the orbital distance is indicated by $r$, which we have assigned to the planetary radius.} The matrix is simply the 2D rotation matrix for the angle $\lambda$.

In our model, we were at this point able to calculate for each observation whether or not parts of the rotating stellar disc are blocked from view. We performed this calculation for each observation and set the flux of pixels covered by the planet to zero when we integrated over the visible stellar surface to obtain a model of the stellar line at particular phases of the transit. The lines were then convolved with a Gaussian whose width is given by the quadrature sum of $\xi$ and $\sigma_{\rm PSF}$. Finally the model CCFs were shifted in velocity space according to a Keplarian model. "Phase smearing" occurs for  integration times that are sufficiently long for the planets movement over the stellar disc to be $\gtrsim r/R$. In such cases, the data should be compared to models that are integrated over such time intervals to emulate phase smearing. However, for the current set of observations, the exposure-time-to-transit-duration ratio is $\approx0.02$, which is less than $r/R\approx0.07$. We therefore do not take this into account here.
  
Now we present our comparison of the model CCFs with the observed CCFs. As in \cite{art:albrecht2013}, we also tried to mitigate the effect that the changes in S/N throughout the night might have on the normalisation of the spectra and resulting CCFs. Such changes might lead to slightly different S/Ns in the CCFs and therefore slightly different CCF heights as well as small overall changes in the CCF baseline. We assigned three parameters to each CCF: an intensity offset, a slope, and a scalar. These parameters were optimized every time a model was compared to the data. This approach serves to propagate the effects of any potential changes in the normalisation into the confidence intervals of the final system parameters. However, in the case of our particular data set,  these parameters vary by less than $0.01\%$ throughout the night as the transit spectra were taken under good and relatively stable conditions (\sref{sec:data}) by a fibre fed spectrograph. 

\begin{figure*}
        \includegraphics[width=\textwidth]{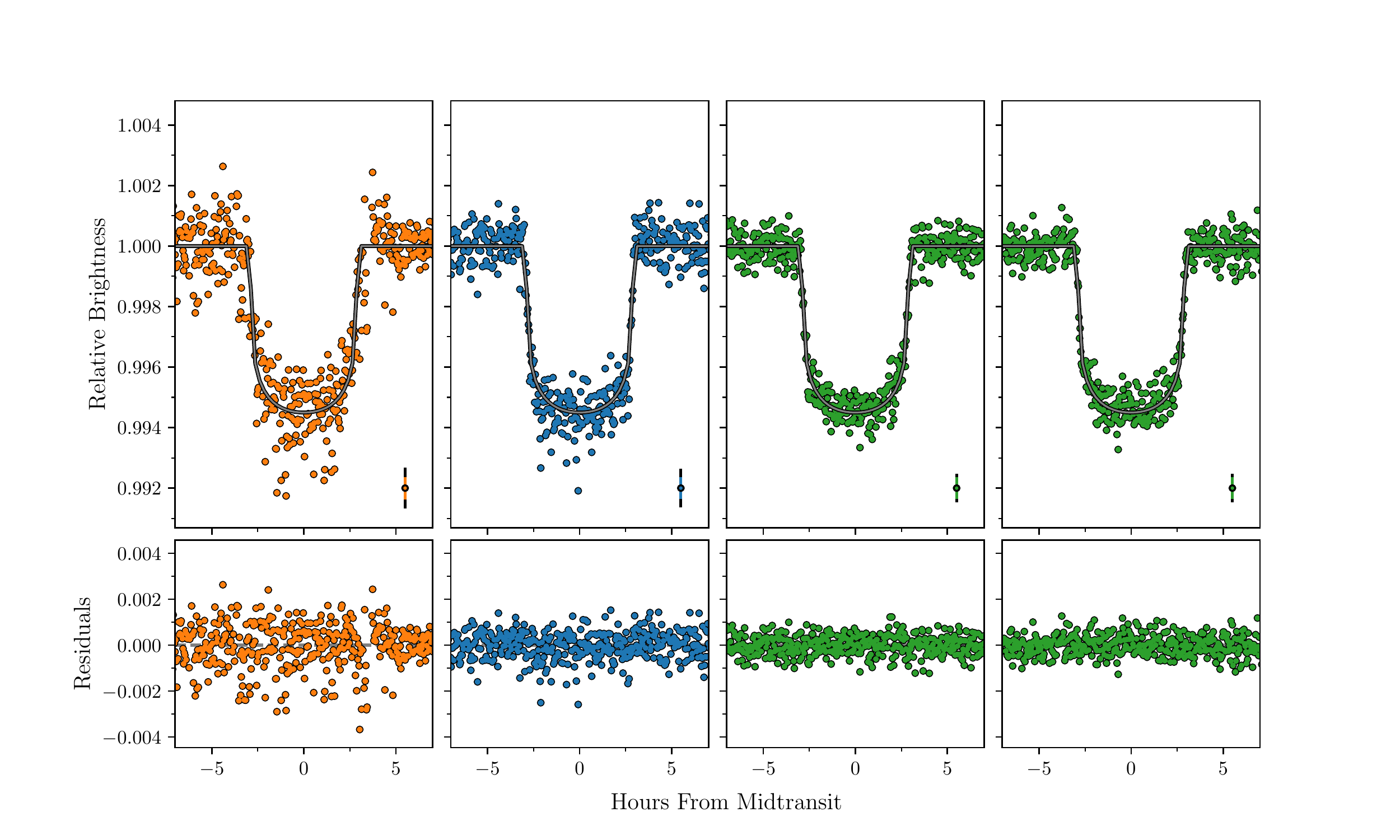}
    \caption{{\bf The four transits of HD~332231~b.} {\bf Top:} Transit data points from the TESS light curve after correction using the GP model seen in \fref{fig:TESS}. The grey line is the best-fitting light curve model. {\bf Bottom:} Residuals from subtracting the best-fitting model from the data. In each top panel, we show an error bar that is representative of the error from this particular selection of data. We added the photometric jitter in quadrature (shown in black) to the nominal error (coloured). }
    \label{fig:lc}
\end{figure*}

\begin{figure*}
    \centering
    \includegraphics[width=\textwidth]{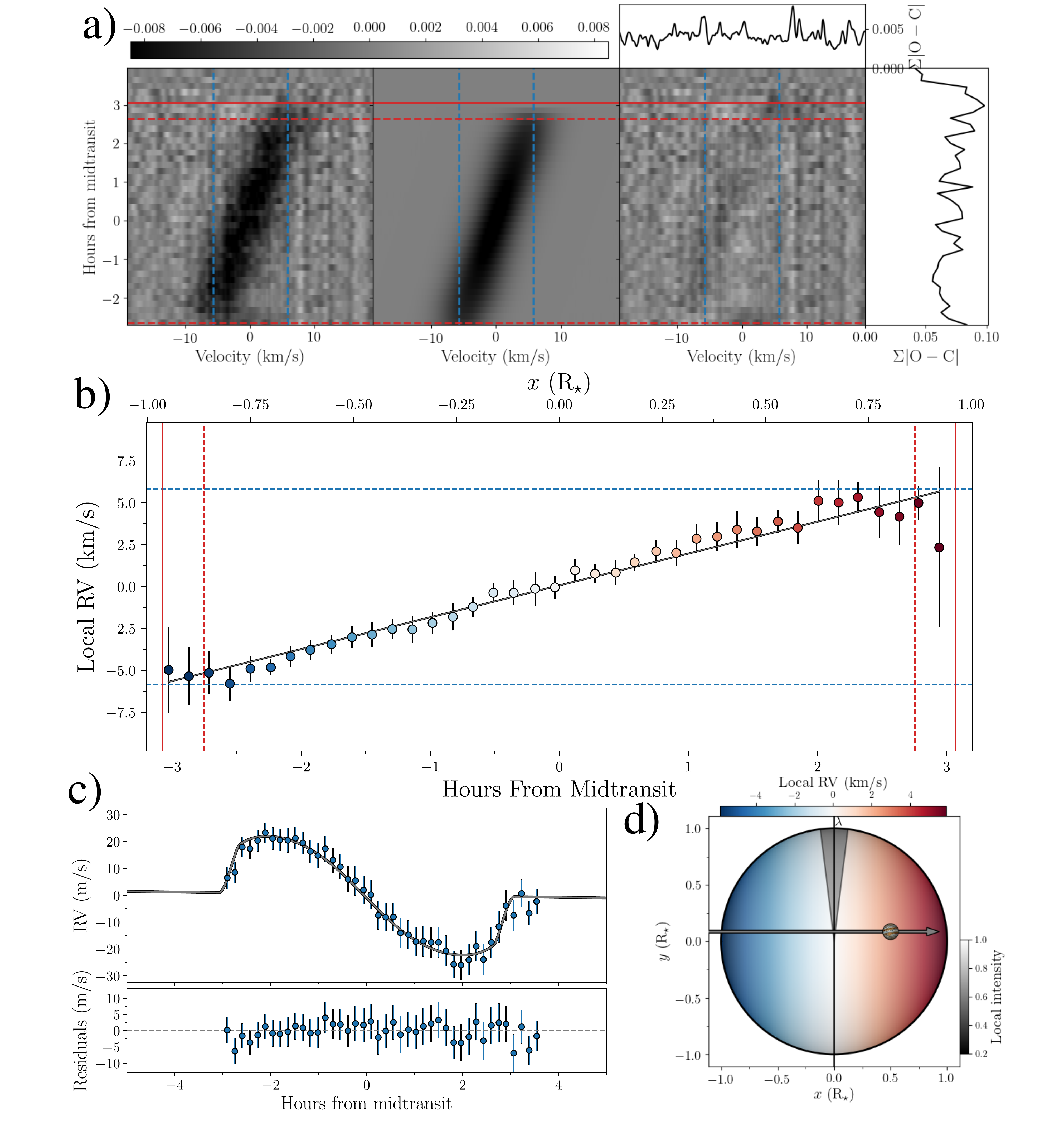}
    \caption{{\bf The three approaches to measuring $\lambda$.} {\bf a)} {\bf Left:} Distortion of the stellar absorption lines as observed with HARPS-N during the transit night. The vertical lines denote $\pm v \sin i$. The horizontal dashed lines are the second and third contact, i.e. the times in between is when the planet is completely within the stellar disc, and the solid line is the fourth contact, i.e. the point where the stellar and planetary discs no longer overlap. {\bf Middle:} Best-fitting model of the distortion of the absorption lines. {\bf Right:} Residuals from subtracting the best-fitting model from the data with the summed values displayed on top and to the right. In all panels, data and model have been shifted into the stellar rest frame, a minor effect. The horizontal colour bar at the top denotes the strength of the signal. {\bf b)} Subplanetary or local RVs created by subtracting the OOT CCF from the in-transit ones and measuring the position of the excess bump caused by the planet. The error bars are colour-coded according to the $v \sin i$ value they should have given their $x$-position in {\bf (d)}, which shows the orientation of the orbit and the projected stellar rotation going from blue ($-v \sin i$) to red ($v \sin i$). As in {\bf (a),} the dashed blue line denotes the value for $v \sin i$, while the solid and dashed red lines mark the contact points. {\bf c)} {\bf Top:}  RVs from HARPS-N used in the fit (shown with blue error bars). The grey line is the best-fitting model of the RM effect. {\bf (c) Bottom:} Residuals from subtracting the best-fitting model from the data. {\bf d)} Geometry of the system with the planet transiting the rotating and limb-darkened star, which is tilted by $\lambda$ with respect to the orbit of the planet marked with the grey arrow. The horizontal colour bar shows the rotation speed at a given longitude, and the grey colour bar shows the relative intensity given going from 1.0 in the centre to around 0.16 at the limb (given the limb-darkening parameters for HARPS-N in \tref{tab:mcmc_res}). Here the intensity overlay has been made transparent to make the rotation colour-coding visible.}
    \label{fig:lambda}
\end{figure*}

In \fref{fig:lc} we show the best-fitting light curve from the MCMC (\sref{sec:mcmc}), while in \fref{fig:lambda} a) we show the HARPS-N data with the best-fitting model as well as the residuals. Here, we subtracted the OOT CCF to highlight the deformation  of the CCFs due to the planetary transit.  The planetary shadow first covers the blueshifted light, and during the second half of the transit covers the redshifted light, consistent with a prograde transit. 

\begin{table*}
    \centering
    \caption{Results from the MCMC runs.}
    \begin{threeparttable}

    \begin{tabular}{l c c c c}
 \toprule 
 Parameter & Prior & RV & {\bf Shadow} & Slope \\ 
 \midrule 
  \multicolumn{5}{c}{Stepping parameters} \\
  \midrule 
$P \rm  \  (days)$ & $\mathcal{U}$ & $18.71205 \pm 0.00001$ & $18.71205 \pm 0.00001$ & $18.71205 \pm 0.00001$ \\ 
$T \rm _{0,b} \ (BTJD)$ & $\mathcal{U}$ & $1729.6814 \pm 0.0004$ & $1729.6814 \pm 0.0004$ & $1729.6814 \pm 0.0004$ \\ 
$\Delta T_\mathrm{0,b} \rm (min.)$ & $\mathcal{U}$ & $20 \pm 3$ & $18 \pm 3$ & $22 \pm 4$ \\ 
$r/R\rm $ & $\mathcal{U}$ & $0.0690 \pm 0.0003$ & $0.0689 \pm 0.0003$ & $0.0690 \pm 0.0003$ \\ 
$a/R \rm $ & $\mathcal{U}$ & $24.3^{+0.4}_{-0.3}$ & $24.4^{+0.4}_{-0.3}$ & $24.4^{+0.4}_{-0.3}$ \\ 
$\cos i_\mathrm{o} $ & $\mathcal{U}$ & $0.006^{+0.003}_{-0.005}$ & $0.0039^{+0.0018}_{-0.0039}$ & $0.0036^{+0.0015}_{-0.0035}$ \\ 
$\lambda \rm  \ (^\circ)$ & $\mathcal{U}$ & $-1 \pm 12$ & $-2 \pm 6$ & $0 \pm 7$ \\ 
$v \sin i \ \rm (km/s)$ & $\mathcal{U}$ & $5.64 \pm 0.14$ & $5.63 \pm 0.11$ & $5.89^{+0.12}_{-0.13}$ \\ 
$\zeta \ \rm (km/s)$ & $\mathcal{N}$(4.46,1.0) & $4.8 \pm 0.4$ & $4.7 \pm 0.3$ & $4.7 \pm 0.2$ \\
$\xi \ \rm (km/s)$ & $\mathcal{N}$(2.0,1.0) & $2.7^{+0.2}_{-0.3}$ & $2.71^{+0.17}_{-0.19}$ & $2.40 \pm 0.17$ \\
$K  \rm \ (m/s)$ & $\mathcal{U}$ & $17.5^{+1.1}_{-1.2}$ & $17.5 \pm 1.1$ & $17.5 \pm 1.1$ \\ 
$\sqrt{e} \ \cos \omega$ & $\mathcal{U}$ & $0.11^{+0.12}_{-0.09}$ & $0.12^{+0.12}_{-0.08}$ & $0.12^{+0.12}_{-0.08}$ \\ 
$\sqrt{e} \ \sin \omega$ & $\mathcal{U}$ & $0.08^{+0.03}_{-0.07}$ & $0.08^{+0.04}_{-0.07}$ & $0.09^{+0.05}_{-0.06}$ \\ 
$\gamma_\mathrm{HARPS-N} \ \rm (m/s)$ & $\mathcal{U}$ & $-23395.1 \pm 1.1$ & $-23346^{+12}_{-11}$ & $-23344 \pm 12$ \\ 
$\rm \sigma_{HARPS-N} \ (m/s)$ & $\mathcal{U}$ & $0.6^{+0.3}_{-0.6}$ & - & - \\ 
$\gamma_\mathrm{Levy} \ \rm (m/s)$ & $\mathcal{U}$ & $1.0 \pm 1.1$ & $1.1 \pm 1.1$ & $1.1^{+1.2}_{-1.0}$ \\ 
$\rm \sigma_{Levy} \ (m/s)$ & $\mathcal{U}$ & $8.7^{+0.8}_{-0.9}$ & $8.7^{+0.8}_{-1.0}$ & $8.7^{+0.8}_{-1.0}$ \\ 
$\gamma_\mathrm{HIRES} \ \rm (m/s)$ & $\mathcal{U}$ & $-1.6^{+1.0}_{-1.1}$ & $-1.6 \pm 1.0$ & $-1.6^{+1.0}_{-1.1}$ \\ 
$\rm \sigma_{HIRES} \ (m/s)$ & $\mathcal{U}$ & $3.4^{+0.9}_{-1.0}$ & $3.4^{+0.8}_{-1.1}$ & $3.4^{+0.9}_{-1.0}$ \\ 
$\gamma_\mathrm{SONG} \ \rm (m/s)$ & $\mathcal{U}$ & $2 \pm 6$ & $2^{+5}_{-6}$ & $2 \pm 6$ \\ 
$\rm \sigma_{SONG} \ (m/s)$ & $\mathcal{U}$ & $18 \pm 5$ & $18^{+4}_{-6}$ & $18^{+4}_{-6}$ \\ 
$q_1 + q_2: \rm  TESS$ & $\mathcal{N}$(0.542,0.1) & $0.55 \pm 0.03$ & $0.56 \pm 0.03$ & $0.55 \pm 0.03$ \\ 
$q_1 + q_2: \rm  HARPS$-$\rm N$ & $\mathcal{N}$(0.7114,0.1) & $0.78 \pm 0.08$ & $0.78 \pm 0.08$ & $0.88^{+0.09}_{-0.10}$ \\ 
$\rm \log \sigma_\mathrm{Sector \ 15}$ & $\mathcal{U}$ & $-7.571^{+0.008}_{-0.007}$ & $-7.571 \pm 0.008$ & $-7.571 \pm 0.008$ \\ 
$\rm \log \tau_\mathrm{Sector \ 15} \ (\log days)$ & $\mathcal{U}$ & $-9.13^{+0.11}_{-0.12}$ & $-9.12^{+0.10}_{-0.12}$ & $-9.13^{+0.11}_{-0.12}$ \\ 
$\rm \log A_\mathrm{Sector \ 15}$ & $\mathcal{U}$ & $-1.0 \pm 0.3$ & $-1.0 \pm 0.3$ & $-1.0 \pm 0.3$ \\ 
$\rm \log \sigma_{Sector \ 14}$ & $\mathcal{U}$ & $-7.502 \pm 0.008$ & $-7.502 \pm 0.008$ & $-7.503 \pm 0.008$ \\ 
$\rm \log \tau_{Sector \ 14} \ (\log days)$ & $\mathcal{U}$ & $-7.92 \pm 0.04$ & $-7.91 \pm 0.04$ & $-7.92 \pm 0.04$ \\ 
$\rm \log A_{Sector \ 14}$ & $\mathcal{U}$ & $-2.68 \pm 0.11$ & $-2.68^{+0.10}_{-0.11}$ & $-2.68^{+0.10}_{-0.11}$ \\ 
$\rm \log \sigma_{Sector \ 41}$ & $\mathcal{U}$ & $-8.127 \pm 0.012$ & $-8.127 \pm 0.012$ & $-8.127^{+0.012}_{-0.013}$ \\ 
$\rm \log \tau_{Sector \ 41} \ (\log days)$ & $\mathcal{U}$ & $-8.46^{+0.12}_{-0.15}$ & $-8.46^{+0.13}_{-0.14}$ & $-8.46^{+0.12}_{-0.15}$ \\ 
$\rm \log A_{Sector \ 41}$ & $\mathcal{U}$ & $-0.23^{+0.16}_{-0.15}$ & $-0.23^{+0.15}_{-0.16}$ & $-0.23^{+0.15}_{-0.16}$ \\ 

\midrule 
\multicolumn{5}{c}{Derived parameters} \\
\midrule 
$i \rm _{o}  \ (^\circ)$ & & $89.67^{+0.29}_{-0.15}$ & $89.78^{+0.22}_{-0.10}$ & $89.80^{+0.20}_{-0.09}$ \\ 
$b \rm $ & & $0.14^{+0.06}_{-0.12}$ & $0.10^{+0.04}_{-0.10}$ & $0.09^{+0.04}_{-0.09}$ \\ 
$e \rm $ & & $0.026^{+0.011}_{-0.024}$ & $0.029^{+0.014}_{-0.024}$ & $0.029^{+0.016}_{-0.022}$ \\ 
$\omega \rm  \ (^\circ)$ & & $40^{+26}_{-40}$ & $39^{+22}_{-39}$ & $39^{+22}_{-39}$ \\ 
$r \ \rm (R_{Jupiter})$ & & $0.857 \pm 0.016$ & $0.856 \pm 0.016$ & $0.857 \pm 0.016$ \\
$q_1: \ \rm  HARPS$-$\rm N$ & & $0.55^{+0.04}_{-0.04}$ & $0.54^{+0.04}_{-0.04}$ & $0.60^{+0.05}_{-0.05}$ \\ 
$q_2: \ \rm  HARPS$-$\rm N$ & & $0.23^{+0.04}_{-0.04}$ & $0.23^{+0.04}_{-0.04}$ & $0.28^{+0.05}_{-0.05}$ \\ 
$q_1: \ \rm  TESS$ & & $0.258^{+0.014}_{-0.014}$ & $0.261^{+0.014}_{-0.014}$ & $0.258^{+0.014}_{-0.014}$ \\ 
$q_2: \ \rm  TESS$ & & $0.294^{+0.014}_{-0.014}$ & $0.297^{+0.014}_{-0.014}$ & $0.294^{+0.014}_{-0.014}$ \\ 

 \bottomrule 
  \end{tabular}
\begin{tablenotes}
    \item Results from the MCMCs using the RVs, the planet shadow, and the local subplanetary velocity. The value is the median of the samples and the upper and lower uncertainties are estimated from the highest posterior density at a confidence level of 0.68. We denote the uniform priors as $\mathcal{U}$ and the Gaussian priors with a mean, $\mu$, and width, $\sigma$, as $\mathcal{N}(\mu,\sigma)$. We chose the parameters obtained from analysing the deformation in the lines (the planetary shadow)  as our final parameters. $T_0$ is given in TESS Barycentric Julian Date (BTJD; $\rm BJD - 2457000$).
\end{tablenotes}
    \label{tab:mcmc_res}
    \end{threeparttable}
\end{table*}

\subsection{Subplanetary velocity}
\label{sec:subpl}
A second approach to measuring $\lambda$ is to determine the subplanetary velocity, $v_{\rm p}$, for each observation and then use a simple geometric model to determine $\lambda$ from the $v_{\rm p}$ measurements \citep[e.g.][]{Cegla+2016,Hoeijmakers+2020}. The results of this method do not depend directly on surface velocity fields. A dependence does remain as the OOT CCFs supply information on $v \sin i$, and that $v \sin i$ measurement does depend on the surface velocity fields. This additional information on $v \sin i$ is particularly important for our system as the impact parameter is close to zero \citep[see][for a discussion on this dependency]{art:albrecht2011}.

The subplanetary velocities are obtained in the following way: the OOT CCF, shifted to the appropriate velocities, is subtracted from the in-transit CCFs. This isolates the distortion in the stellar lines, i.e. the planet shadow. The central subplanetary velocity is then measured by fitting a Gaussian to the distortions during transit, where we only searched for a distortion of the CCFs inside the interval $\pm 2 \times v \sin i$. We used the uncertainties derived from the co-variance matrix of a Levenberg-Marquardt fit and then further increased them in quadrature so $\chi^{2}_\nu \approx 1$. We extracted $v_{\rm p}$ for each set of system parameters afresh out of concern that any error we make in isolating the planet shadow, by for example not using the proper orbital velocity, might lead to a systematic error in the measured $v_{\rm p}$. However, we tested that even for line parameters significantly outside our confidence intervals, the subtraction of the overall line and subsequent determination of $v_{\rm p}$ does not change $v_{\rm p}$ outside its uncertainty interval. The extracted velocities and their uncertainties are given in \tref{tab:RVs} and can be seen in \fref{fig:lambda} b). 

From Eq.~\eqref{equ:v_p} it is clear that $v_{\rm p}$ only depends on the $x$-coordinate of the planet and should progress linearly with time. Therefore, in our model we can calculate $v_{\rm p}$ with a first-order polynomial, where extremes occur at ingress $V_{\rm ingress}$ and $V_{\rm egress}$ and both can be taken as positive values. The offset and amplitude of the line are given by \citep[][Albrecht et al.~in prep]{art:albrecht2011},
\begin{equation}
    \begin{split}
    V_{\rm egress} - V_{\rm ingress}  & = 2 \times (v \sin i) \sin\lambda \times b \, ,  \\
    V_{\rm egress} + V_{\rm ingress} & = 2 \times (v \sin i) \cos\lambda \times \sqrt{1-b^2} \, .
    \end{split}
\end{equation}
For this particular system with $b\approx0$, good alignment would be indicated with $V_{\rm egress} + V_{\rm ingress}\approx 2 \times (v \sin i) \approx 11$~km\,s$^{-1}$. 

\subsection{Radial velocities}
The distortions of the spectral lines as seen in \fref{fig:lambda} (a) lead to anomalous RVs observed during transit, displayed in \fref{fig:lambda} (c). A first-order estimate of the anomalous stellar RVs can be obtained from
\begin{equation}
{\rm RV_{RM}}(t) \approx -\left( \frac{r}{R} \right)^2 v_{\rm p}(t) \, .
\end{equation}
The RV$_{\rm RM}$ measurements relate to $v_p$ and the radius ratio of the transiting to the occulted object. The sign change occurs as the subplanetary light is blocked from view. Any particular RV$_{\rm RM}$ is further modified by the stellar limb darkening at the subplanetary point on the stellar disc, Eq.~\eqref{equ:limb_darkening}, and during ingress and egress by the ratio of the planetary disc in front of the star. This can be seen by comparing \fref{fig:lambda} (b) to \fref{fig:lambda} (c). For our RV model, we used the algorithm by \citet{art:hirano2011} which also includes the effect of instrumental and stellar broadening. For the other two approaches,  here we also include a Keplerian RV model. 
 
\subsection{Comparison of data and model }
\label{sec:mcmc}
 
To extract confidence intervals for the system parameters, we used MCMCs. We define our likelihood function as,
\begin{equation}
    \label{equ:likelihood}
    \log \mathcal{L} =-0.5 \sum_{i=1}^{N} \left [ \frac{(O_i - C_i)^2}{\sigma_i^2} + \log 2 \pi \sigma_i^2 \right] + \sum_{j=1}^{M} \log \mathcal{P}_{j}\, ,
\end{equation}
where $N$ indicates the total number of data points from photometry, archival RVs, and the spectroscopic data obtained during the transit night; $C_i$ represents the model corresponding to the observed data point $O_i$; $\sigma_i$ is the uncertainty for the $i$th data point, where we add a jitter term in quadrature and a penalty in the likelihood for the RVs as well as photometry; and $\mathcal{P}_j$ denotes the prior for the $j$th parameter listed in \tref{tab:mcmc_res}. 

We did not step directly in the limb-darkening coefficients; rather we were stepping in the sum of the two, that is $q_1 + q_2$, where we applied a Gaussian prior with a width of 0.1. The difference, that is $q_1 - q_2$, was kept fixed during the sampling. Furthermore, our MCMC was stepping in $\sqrt{e} \cos \omega $, $\sqrt{e} \sin \omega $, and $\cos i$. We used \texttt{emcee} \citep{art:foremanmackey2013} to carry out the   MCMC sampling of the posteriors. To ensure that our MCMC runs converged, we invoked the rank-normalised $\hat{R}$ diagnostic test \citep{art:vehtari2019} using the \texttt{rhat} module implemented in \texttt{ArviZ} \citep{misc:arviz2019}. Our results for all three approaches are given in \tref{tab:mcmc_res}, and all clearly suggest that the stellar spin axis and the orbital axis of the planet are aligned. 

\section{Results}
\label{sec:results}

The additional TESS photometry from Sector 41 allows us to improve the planet-to-star radius ratio from that of the discovery paper, shrinking the uncertainty by $\sim 33\%$. We find $r/R=0.0689 \pm 0.0003$. This is because (i) now four instead of two transits have been observed and (ii) Sector 41 photometry has a lower scatter. Using our new value for $r/R$ with the stellar radius given in \citet[][(see our \tref{tab:toi-1456_para}),]{art:dalba2020} we find $r=0.857\pm0.016$~R$_\mathrm{Jupiter}$, which is consistent with the discovery value of $0.867^{+0.027}_{-0.025}$~R$_{\rm Jupiter}$. 

There is a discrepancy in the systemic velocity for the HARPS-N data, $\gamma_{\mathrm{HARPS}-\mathrm{N}}$, between the three approaches; see \tref{tab:mcmc_res}. The shadow and slope approaches give consistent results with each other but not with the RV method. While we have not further investigated this here, we suspect that this disagreement is caused by the simple stellar line model we employ in the shadow and slope approaches. In particular, we do not model convective blueshift as done by \citet{AlbrechtWinnJohnson+2012} using the approach by \citet{shporer2011}. Therefore, our model of the line shape is fully symmetric, which is the opposite of what we expect for actual stellar absorption lines. While convective blueshift might influence all three approaches, the difference in $\gamma$ probably comes about as the line and slope method fits the line itself {and} the velocity position of the distortion $v_p$, while the RV method only fits the RVs. Given the fast rotation of the star and the quality of our data, convective blueshift might have influenced the result of $\gamma$  significantly, but not the result for $\lambda$. Ideally, magnetohydrodynamics simulations of the stellar photosphere should be used to model the lines \citep{Cegla+2016,art:dravins2017}. We further note that the values for $v\sin i$ and $q_1+q_2 $~derived from HARPS-N data come out slightly larger for the slope compared to the values for the RV and shadow runs. These two parameters are correlated in that, for a stronger limb darkening, $v \sin i$ needs to be larger as well to fit the OOT. 

\begin{figure*}
    \centering
    \includegraphics[width=\textwidth]{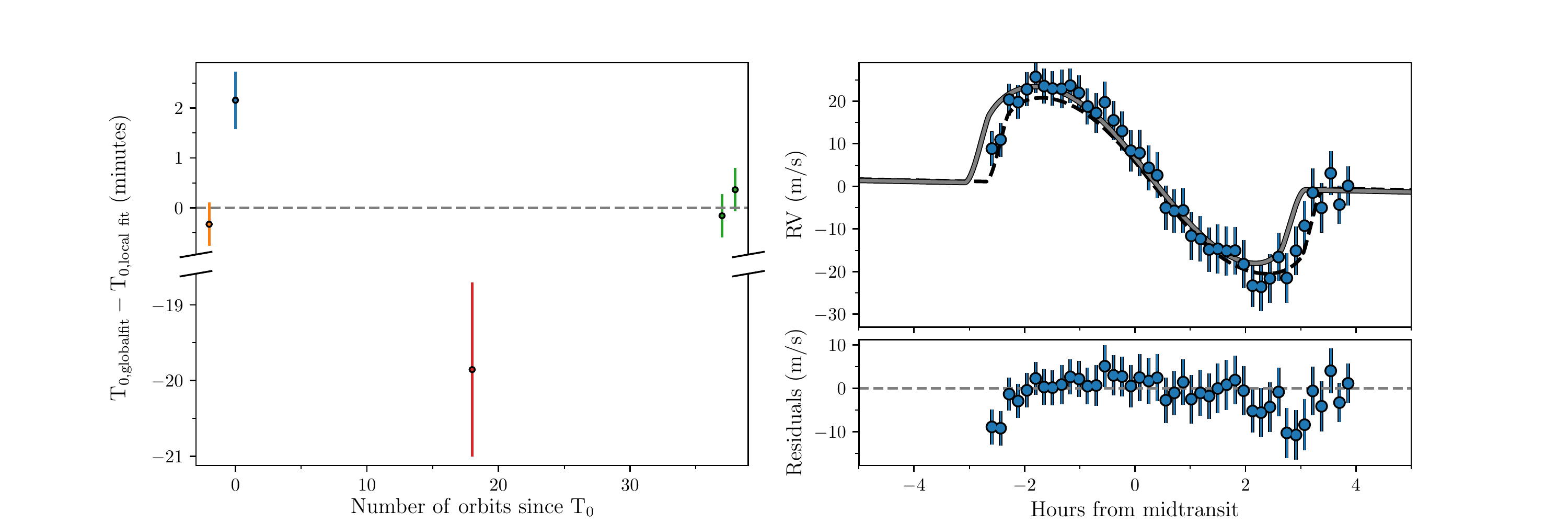}
    \caption{{\bf Transit timing differences.} {\bf Left:} Timing differences between the predictions of transit midpoints derived from linear ephemeris obtained via a global fit to all photometric data and transit midpoints measured for each of the transits individually (including the spectroscopic transit), i.e., $T_0$ from \tref{tab:mcmc_res}. We also show a local fit, where $T_0$ is free to vary for each of the specific transits. The transits from TESS Sectors 14, 15, and 41 are plotted in orange, blue, and green, respectively. The transit observed with HARPS-N is plotted with a red symbol. The spectroscopic transit is offset by several $\sigma$ from its expected timing. {\bf Right:} Similar to \fref{fig:lambda} panel (c), but now assuming linear ephemeris resulting in a misaligned ($\lambda=-31 \pm 6$~$^\circ$) orbit shown in grey. Clearly, this model does not fit the data as well as the dashed line, which is the model in panel (c) of \fref{fig:lambda}. The lower panel displays the residuals between the linear ephemeris model and data. There are large systematic differences between data and model, specifically near ingress and egress, as expected if there is a timing offset. }
    \label{fig:timing}
\end{figure*}

Before the recent release of data from TESS Sector 41, we found in our initial runs that the three approaches of measuring $\lambda$ lead to  inconsistent results. Specifically, the subplanetary velocity approach found alignment ($\lambda=-7 \pm 8^{\circ}$), the shadow analysis indicated moderate misalignment ($\lambda=-16 \pm 4^{\circ}$), and the RV$_{\rm RM}$ measurements suggested a very significant misalignment, $\lambda=-31 \pm 6^{\circ}$. 

Including TESS Sector 41 data, we now have photometry obtained during July and August 2019 and July and August 2021 bracketing our spectroscopic transit observations from the \transitnight. From this, it now appears that the midpoint of the transit we observed with HARPS-N ($T_0=1729.6811$ in BTJD; TESS Barycentric Julian Date; $\rm BJD - 2457000$) is shifted by $\sim20$~min relative to the expected value from the linear ephemeris as derived from TESS photometry alone. The left panel of \fref{fig:timing}  shows the deviation of the measured transit midpoints from these particular ephemerides. 

This apparent mismatch in the mid-transit time has a more significant influence on the result obtained from the RVs than that from the shadow and slope. This is because the latter two methods are less governed by ingress and egress data. A shift by a few minutes will lead to a large difference between the RV$_{\rm RM}$ model and data (\fref{fig:timing}). Such a difference is largest for ingress and egress data. Such data contribute less to the results for the shadow or slope methods and more importantly a shift in timing between model and data can be absorbed into the systemic velocity without significantly influencing the result for $\lambda$. RV$_{\rm RM}$ data ---if pre-ingress and/or post-egress data have been obtained--- do not allow for such a shift.  

\begin{figure}
    \centering
    \includegraphics[width=\columnwidth]{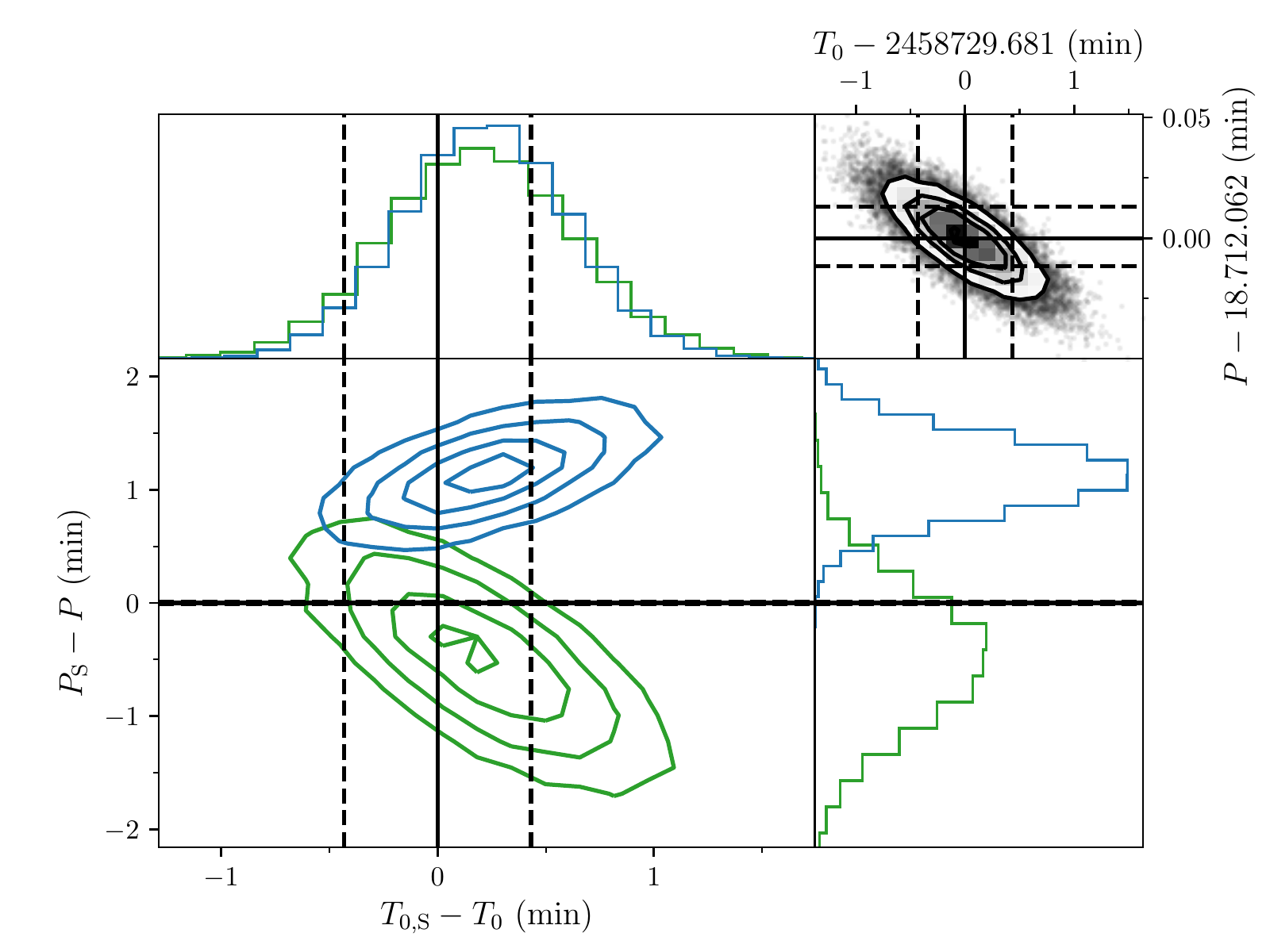}
    \caption{{\bf Correlation between $P$ and $T_0$.} Fit varying $P$ and $T_0$, but otherwise fixing the parameters from the RV run in \tref{tab:mcmc_res}. In blue we only fit the transits in Sectors 14 and 15, and in green we only fit the transits from Sector 41. We have subtracted $P$ and $T_0$ from \tref{tab:mcmc_res} from the results. Displayed in the top right corner are the posteriors for these, and their values and confidence intervals are shown as black lines in both correlation plots. Furthermore, we used this period to shift $T_0$ from Sector 41, i.e. subtracting $37 \times P$. The confidence intervals for the period, translating to different mid-transit points for different epochs, are only marginally consistent, indicating the possibility of TTVs.}
    \label{fig:ephemeris}
\end{figure}

We investigated this further by plotting the posteriors for $P$ and $T_0$ as obtained from photometry only. We derived these using data from Sectors 14 and 15 only, from Sector 41 only, and from all sectors combined. The results can be seen in \fref{fig:ephemeris}. The results from the different sectors appear to be only marginally consistent.  
 Other parameters showed inconsistencies in addition to $\lambda$. The impact parameter was found to be $0.23 \pm 0.05$. This value is inconsistent with the value obtained from photometry alone.  

Given these considerations we conclude that a model employing a linear ephemeris is not adequate for modelling the system. However, we do not have enough data to constrain a physical model predicting such transit timing variations (TTVs) for the observed spectroscopic transit. We therefore decided to introduce an additional parameter $\Delta T_0$. This parameter allows the mid-transit time of the specific transit observed with HARPS-N to float freely relative to the prediction from the linear ephemeris, now only determined from the TESS photometry. Adding this additional parameter will reduce the precision of our final result in $\lambda$ as we ask our spectroscopic data to constrain an additional parameter, which would have otherwise been constrained by the photometry. However, this is a conservative choice as any possible TTVs will now not bias our result for the projected obliquity. 

\begin{figure*}
    \centering
    \includegraphics[width=0.49\textwidth]{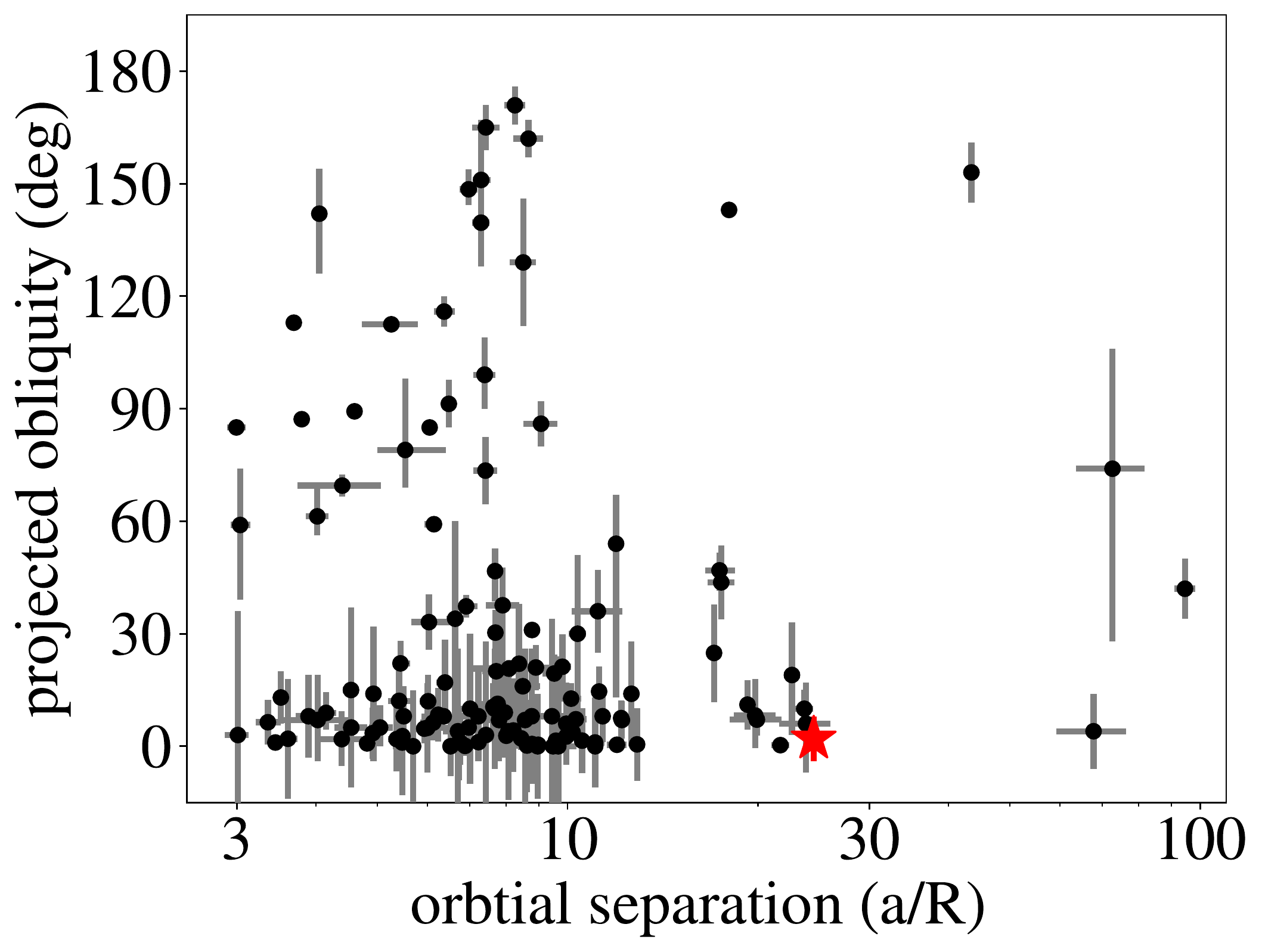}
    \includegraphics[width=0.49\textwidth]{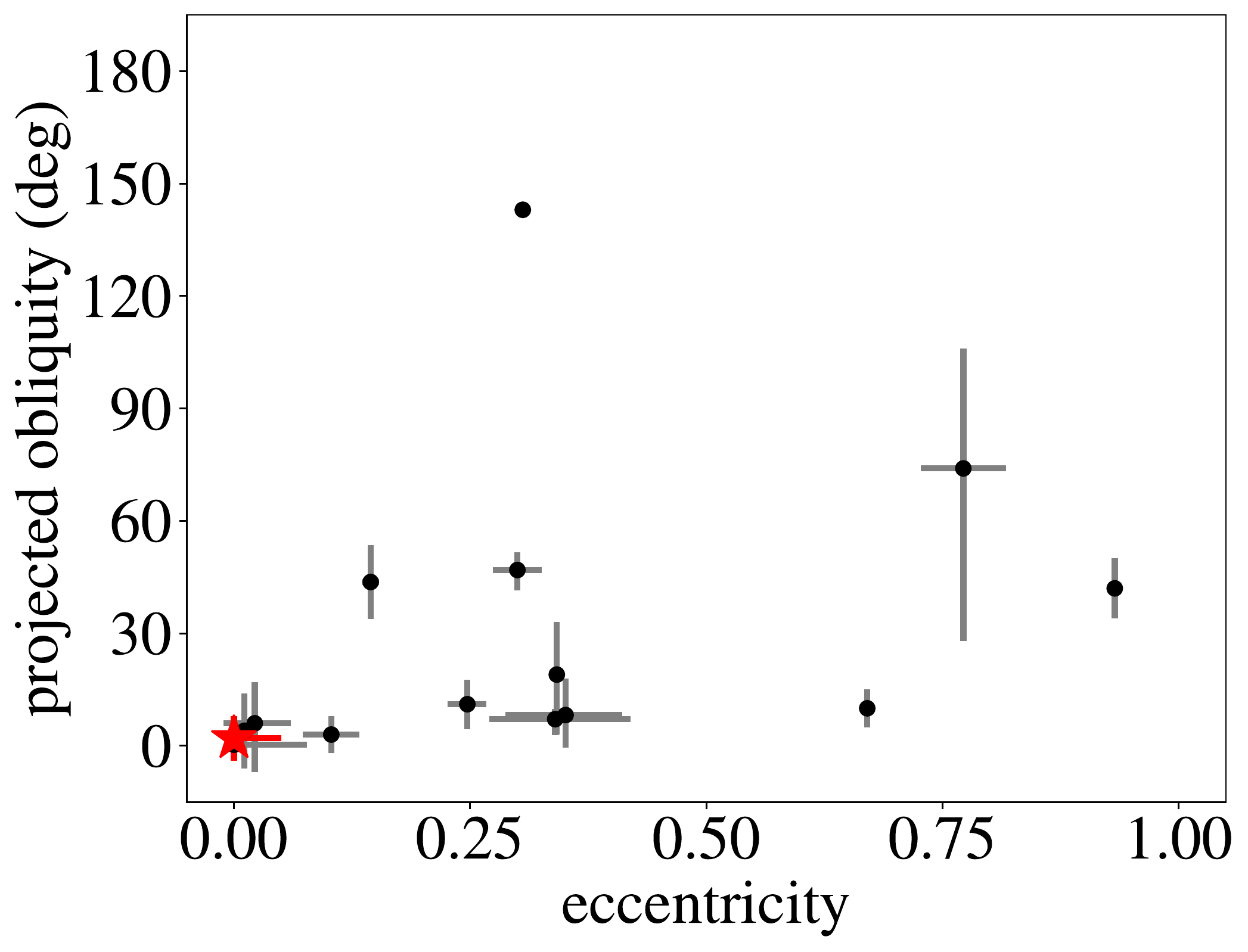}
    \caption{{\bf Comparing HD~332231 to the literature.} {\bf Left:} Projected obliquities versus orbital separations {\bf Right:} Projected obliquities versus orbital eccentricities.  We only include planets   more massive than $0.2$~M$_{\rm Jupiter}$. In the right panel, we further only plot planets on large ($a/R > 15$) orbits and with $\sigma_{\rm ecc}<0.1$. \host is marked by a red star; the system belongs to the class of systems with low orbital misalignment, and low orbital eccentricity.}
    \label{fig:ar_ecc}
\end{figure*}

Implementing this change to our model, the goodness of the model to data comparison improved, as measured via $\chi^2$ for all approaches we employed to determine $\lambda$. The most noticeable improvement occurred for the $\chi^2$ of the RM RV approach, where $\chi^2$ decreased by factor of $\approx2.8$. More importantly, the functional shape of the best fitting RM RV model is a much better fit to the anomalous RVs obtained during transit. This is clearly seen when comparing the residuals of the right panel in \fref{fig:timing} to \fref{fig:lambda} (c). Finally, the impact parameters obtained from all three methods now agree, as expected, with the value obtained from photometry alone $b=0.10^{+0.04}_{-0.10}$. We therefore regard this model, including a time-shift, as the most appropriate. 

We have a remaining concern in regards to the RV data. As can be seen from the lower panel in \fref{fig:lambda} (c), almost all RV data points from the transit night agree within their 1~$\sigma$ uncertainties with the model. This suggests that the uncertainty intervals provided by the DRS\footnote{The "HIERARCH TNG DRS DVRMS" entry in the FITS header.} pipeline are too conservative. There are six effective parameters that are (also) controlled by the HARPS-N RVs ($\Delta T_0$, $\gamma_{\rm HARPS-N}$, $\lambda$, $v \sin i$, $q_1+q_2$, $\zeta$, and $\xi$, of which the last four are further constrained by priors and the OOT CCF). There are $42$ HARPS-N RV measurements resulting in $36$ degrees of freedom. We find a $\chi^2$ of $10.07$ and a reduced $\chi^2$ of $0.28$. This would suggest that the uncertainty interval we obtain for the projected obliquity from the fit to the RVs might be overestimated. Indeed, the interval is about twice the size as for the other two methods. Another indication that the RV confidence interval might be overestimated comes from a comparison to the other two approaches. Given the moderate projected rotation speed of the star, the uncertainty intervals between these different methods should not differ by a factor $\sim2,$ as the shadow and slope methods would benefit more from the host having a larger $v \sin i$. 

In summary, we find that our results for $\lambda$ ---using the three different methods on the same data--- are consistent with each other  ($\lambda=-1\pm12$~$^\circ$, $\lambda=-2\pm6$~$^\circ$, $\lambda=0\pm7$~$^\circ$) as expected; see \tref{tab:mcmc_res}. Given the uncertainty interval from the fit to the anomalous RVs (which is probably too large as a result of overly conservative RV uncertainties), we use the projected obliquity value as derived from the fit to the shadow as a final parameter, meaning $\lambda=-2\pm6$~$^\circ$ is our final value for the projected obliquity.

\section{Discussion}
\label{sec:discussion}

While finalising this manuscript we became aware of a recent measurement of the projected obliquity in this system by \citet{art:sedaghati2021}. These authors observed \host during a transit night in October 2020, employing the CARMENES high-resolution spectrograph \citep{art:quirrenbach2014} installed at the 3.5~m telescope at the Calar Alto Observatory, Spain. This team finds a projected obliquity of $\lambda=-42.0^{+11.3}_{-10.6}$~$^\circ$. This value is not in agreement with our final result, $\lambda=-2\pm6$~$^\circ$. We suspect that there might be at least two reasons for the apparent disagreement between these two measurements. Firstly, these latter authors relied on priors solely derived from the Sector 14 and 15 TESS photometry by \citet{art:dalba2020}, and their spectroscopic transit observations were obtained about a year after these TESS observations (about two months after our spectroscopic transit observations). Therefore, their assumption of linear ephemeris might be as erroneous as our initial assumption of linear ephemeris. Indeed, the result of their analysis of the anomalous RVs observed during transit is similar to our initial result ($\lambda=-31 \pm 6^{\circ}$) using the linear ephemeris derived from the TESS photometry.

A second potential reason for the mismatch might be connected to their result for the projected stellar rotation speed. They find $v\sin i = 16.3^{+6.9}_{-4.4}$~km~s$^{-1}$. This is significantly higher than the three values ($5.3\pm 1.0$~km~s$^{-1}$, $5.4\pm 1.0$~km~s$^{-1}$, and $7.0\pm 0.5$~km~s$^{-1}$) reported in the discovery paper \citep{art:dalba2020}. That value is also difficult to reconcile with the width of the CCFs we observed; see \fref{fig:ccf}. Therefore, it is also significantly higher than what we find in our overall analysis. As discussed in \citet{art:albrecht2011} for low-impact transits such as this one, a prior on $v \sin i$ can have significant influence on the derived value for $\lambda$. It would be interesting to investigate whether using a lower $v\sin i$ value, as well as allowing for a shift in transit mid-time, would not only lead to a consistent result for $\lambda$, but also to a similar result for the specific transit mid-time, as the spectroscopic transits are only two months apart.

With the data at hand, we cannot determine the cause of the departure from linear ephemeris  with certainty. One possibility would be a second planet whose gravitational influence perturbs the orbit of planet b from a purely Keplerian orbit. The presence and parameters of such a potential third body may be investigated by additional RV monitoring to detect long-term RV drifts, and ground-based transit observations to measure TTVs. Also, in August 2022, the system will be observed again in TESS Sector 55. Additionally, upcoming {\it Gaia} \citep{art:gaia2016} data releases should be able to further clarify the nature of this potential body by detecting or giving upper limits to the reflex motion of the central star caused by such a potential outer body.

Regarding the additional RV monitoring, \citet{art:dalba2020} (Section 4.3) investigated the possibility of a linear trend in their RVs. These authors found that a model including a first-order acceleration term is indistinguishable from a model without an acceleration parameter. As our HARPS-N transit observations do not add to the baseline given the RV offset between the spectrographs, the conclusion remains that we are currently not in a position to expand further on this.

Our result for the projected obliquity, $\lambda=-2\pm 6^\circ$, suggests that the obliquity $\psi$ is also low, and that the system is spin-orbit aligned. The finding that $\psi \sim \lambda$ is supported by the effective temperature of $\sim 6100$~K and the projected stellar rotation of $\sim5.7$~km~s$^{-1}$, which according to \citet{art:louden2021} is consistent with a stellar inclination of close to $90^\circ$.

Given the $a/R$ of about $24.5,$ this system might not have been influenced by tides. Therefore, the alignment we are seeing might be primordial. This, together with the low eccentricity of the orbit ($e = 0.029^{+0.014}_{-0.024}$), fits in a picture that orbital inclinations are often associated with high eccentricities, as illustrated in \fref{fig:ar_ecc}. In this picture, \pl has not undergone high-eccentricity migration. \citet{dawson_murray-clay2013} noticed that eccentric WJs tend to orbit metal rich ([Fe/H]~$>0$) hosts while planets on circular orbits tend to be found around stars with lower metallicities. The solar metallicity measured for \host agrees with this picture.  

\section{Conclusion}
\label{sec:conclusion}

We measured the projected obliquity of the bright F8 dwarf \host using HARPS-N data acquired during transit of the warm giant planet \pl discovered by \citet{art:dalba2020}. We used three different approaches to analyse the RM effect.  We model the planet shadow, the subplanetary velocities, and the anomalous in-transit RVs, obtaining fully consistent results. Our measurement of the projected obliquity $\lambda=-2\pm6$~$^\circ$ is consistent with alignment. Since the discovery of the planet, additional TESS photometry from Sector 41 has become available and we use this here to further refine the system parameters, specifically the planetary radius and linear ephemeris. We find an apparent shift of $\approx 20$~min in the mid-transit time of the transit observed on \transitnight, with HARPS-N at the TNG. This shift is relative to the linear ephemeris obtained from transits observed by TESS Summer 2019 and 2021. This shift might be explained by the presence of a third body in the system and future RV, transit, and astrometric observations should be able to find such a body. While there is a non-zero probability that the system will end up in an aligned configuration through a violent dynamical process, the most probable interpretation of our findings is that the system architecture is the result of the planet having migrated to its current orbit via disc migration or that it was born in situ.

\begin{acknowledgements}
The authors thank the anonymous referee for a helpful review of this work.
The authors thank Paul Dabla for early discussions on TOI-1456.
The authors thank Antonio Magazzu and Rosario Consentino at the TNG as well as Davide Gandolfi for helping with the HARPS-N data.
The authors thank Debra Fischer for attempting to observe an ingress of the transit in HD~332231 with the Lowell Discovery Telescope before it was observed with HARPS-N at the TNG.
Based on observations (programme  ID:  A41/TAC19) made with the Italian Telescopio Nazionale Galileo (TNG) operated on the island of La Palma by the Fundación Galileo Galilei of the INAF (Istituto Nazionale di Astrofisica) at the Spanish Observatorio del Roque de los Muchachos of the Instituto de Astrofisica de Canarias.
Funding for the Stellar Astrophysics Centre is provided by The Danish National Research Foundation (Grant agreement no.: DNRF106).
This paper includes data collected with the TESS mission, obtained from the MAST data archive at the Space Telescope Science Institute (STScI). Funding for the TESS mission is provided by the NASA Explorer Program. STScI is operated by the Association of Universities for Research in Astronomy, Inc., under NASA contract NAS 5–26555.
The numerical results presented in this work were obtained at the Centre for Scientific Computing, Aarhus \url{http://phys.au.dk/forskning/cscaa/}.
This research made use of Astropy,\footnote{http://www.astropy.org} a community-developed core Python package for Astronomy \citep{art:astropy2013,art:astropy2018}. 
This research made use of matplotlib \citep{misc:hunter2007}.
This research made use of TESScut \citep{art:brasseur2019}.
This research made use of astroplan \citep{misc:morris2018}.
This research made use of SciPy \citep{misc:scipy2020}.
This research made use of corner \citep{art:foremanmackey2016}.
\end{acknowledgements}

%
%

\bibliographystyle{aa}
\bibliography{main}

\begin{appendix}
\section{Table of radial and subplanetary velocities}

\begin{table}
\centering
\caption{RVs and subplanetary velocities from HARPS-N.} \label{tab:RVs} 
\begin{threeparttable}
\begin{tabular}{ccc}
\toprule
Epoch & RV$-\gamma_{\mathrm{HARPS}-\mathrm{N,\ RV}}$ & $v_{\rm p}$ \\ 
 BTJD$_{\rm TDB}$ & m\,s$^{-1}$ & km\,s$^{-1}$  \\ 
\midrule
2066.389858 & 9.1 $\pm$ 4.0 & -4.9 $\pm$ 2.6 \\ 
2066.396375 & 11.1 $\pm$ 4.0 & -5.3 $\pm$ 1.7 \\ 
2066.402856 & 20.5 $\pm$ 3.7 & -5.1 $\pm$ 1.3 \\ 
2066.409604 & 20.0 $\pm$ 3.9 & -5.7 $\pm$ 1.0 \\ 
2066.416248 & 23.0 $\pm$ 4.0 & -4.8 $\pm$ 0.8 \\ 
2066.422834 & 25.9 $\pm$ 3.8 & -4.8 $\pm$ 0.5 \\ 
2066.429211 & 23.7 $\pm$ 4.1 & -4.1 $\pm$ 0.6 \\ 
2066.435565 & 23.2 $\pm$ 4.0 & -3.7 $\pm$ 0.6 \\ 
2066.442417 & 23.1 $\pm$ 4.5 & -3.4 $\pm$ 0.6 \\ 
2066.449038 & 23.8 $\pm$ 4.1 & -3.0 $\pm$ 0.6 \\ 
2066.455485 & 22.1 $\pm$ 4.1 & -2.8 $\pm$ 0.7 \\ 
2066.462047 & 19.0 $\pm$ 4.2 & -2.5 $\pm$ 0.6 \\ 
2066.468506 & 17.4 $\pm$ 4.7 & -2.5 $\pm$ 0.8 \\ 
2066.474999 & 19.9 $\pm$ 4.8 & -2.1 $\pm$ 0.7 \\ 
2066.481608 & 15.7 $\pm$ 4.6 & -1.8 $\pm$ 0.8 \\ 
2066.488055 & 13.2 $\pm$ 4.6 & -1.2 $\pm$ 0.6 \\ 
2066.494652 & 8.6 $\pm$ 4.9 & -0.4 $\pm$ 0.6 \\ 
2066.501215 & 8.0 $\pm$ 5.5 & -0.4 $\pm$ 0.7 \\ 
2066.508125 & 4.5 $\pm$ 5.3 & -0.1 $\pm$ 1.0 \\ 
2066.514525 & 2.8 $\pm$ 5.4 & -0.0 $\pm$ 0.7 \\ 
2066.521007 & -4.8 $\pm$ 5.2 & 1.0 $\pm$ 0.6 \\ 
2066.527465 & -5.5 $\pm$ 5.1 & 0.8 $\pm$ 0.5 \\ 
2066.534144 & -5.4 $\pm$ 5.2 & 0.8 $\pm$ 0.7 \\ 
2066.540255 & -11.4 $\pm$ 5.6 & 1.4 $\pm$ 0.5 \\ 
2066.547223 & -12.1 $\pm$ 5.3 & 2.1 $\pm$ 0.7 \\ 
2066.553658 & -14.6 $\pm$ 5.2 & 2.0 $\pm$ 0.8 \\ 
2066.560221 & -14.5 $\pm$ 5.7 & 2.8 $\pm$ 0.9 \\ 
2066.566876 & -14.9 $\pm$ 5.8 & 3.0 $\pm$ 0.9 \\ 
2066.573276 & -14.9 $\pm$ 5.6 & 3.4 $\pm$ 1.1 \\ 
2066.579735 & -18.0 $\pm$ 5.6 & 3.3 $\pm$ 0.8 \\ 
2066.586494 & -23.1 $\pm$ 5.0 & 3.9 $\pm$ 0.7 \\ 
2066.592895 & -23.4 $\pm$ 5.7 & 3.5 $\pm$ 1.0 \\ 
2066.599550 & -21.4 $\pm$ 5.7 & 5.1 $\pm$ 1.2 \\ 
2066.606078 & -16.4 $\pm$ 5.6 & 5.0 $\pm$ 1.4 \\ 
2066.612305 & -21.3 $\pm$ 5.8 & 5.3 $\pm$ 0.9 \\ 
2066.619215 & -14.9 $\pm$ 5.7 & 4.5 $\pm$ 1.6 \\ 
2066.625696 & -9.0 $\pm$ 5.9 & 4.3 $\pm$ 2.6 \\ 
2066.631854 & -1.3 $\pm$ 5.6 & 5.0 $\pm$ 1.0 \\ 
2066.638544 & -4.8 $\pm$ 5.8 & 2.5 $\pm$ 5.2 \\ 
2066.645430 & 3.3 $\pm$ 5.2 & - \\ 
2066.652039 & -4.1 $\pm$ 4.5 & - \\ 
2066.658498 & 0.3 $\pm$ 4.5 & - \\ 
\bottomrule
\end{tabular}
\begin{tablenotes}
    \item The RVs and subplanetary velocities determined from our HARPS-N data. The epoch refers to the flux-weighted midpoint of the observations. From the RVs listed here we have subtracted the systemic velocity, $\gamma_{\mathrm{HARPS}-\mathrm{N}}$, from our MCMC using the RVs (\tref{tab:mcmc_res}).
\end{tablenotes}
\end{threeparttable}
\end{table}

\end{appendix}

\end{document}